\documentclass[prl,twocolumn,showpacs,amsmath,amssymb,superscriptaddress]{revtex4-2}

\usepackage{amsmath,amssymb,amsfonts,bm,color,graphicx,tabularx}

\usepackage[unicode=true,colorlinks=true]{hyperref}

\usepackage[etex=true,export]{adjustbox}
\usepackage{makecell}
\usepackage{multirow}
\usepackage{verbatim}

\hypersetup{linkcolor=blue, citecolor=blue,urlcolor=blue}
\usepackage{tabularx,multirow,array,diagbox}

\newcommand{\beq}{\begin{equation}}
\newcommand{\eeq}{\end{equation}}
\newcommand{\beqn}{\begin{eqnarray}}
\newcommand{\eeqn}{\end{eqnarray}}

\newcommand{\Jt}{\bm{j}}

\newcommand{\qv}{\bm{q}}
\newcommand{\vv}{\mathbf{v}}

\newcommand{\Pg}{P_{g}}
\newcommand{\Rg}{R_{g}}
\newcommand{\Ug}{U_{g}}
\newcommand{\etag}{\eta_{g}}

\newcommand{\cT}{\mathcal{T}}
\newcommand{\PT}{P\cT}

\begin{document}

\title{Phonon Angular Momentum Multipoles and Nonlinear Thermal Responses}

\author{Xun-Jiang Luo}
\email{xjluo@hmfl.ac.cn}
\affiliation{Anhui Province Key Laboratory of Low-Energy Quantum Materials and Devices, High Magnetic Field Laboratory, HFIPS, Chinese Academy of Sciences, Hefei, Anhui 230031, China}
\author{Mingliang Tian}
\email{tianml@hmfl.ac.cn}
\affiliation{Anhui Province Key Laboratory of Low-Energy Quantum Materials and Devices, High Magnetic Field Laboratory, HFIPS, Chinese Academy of Sciences, Hefei, Anhui 230031, China}

\begin{abstract}
Chiral phonons carry angular momentum and couple lattice dynamics to
magnetic and transport phenomena. In magnets, however, a systematic
classification of momentum-space phonon angular momentum (PAM) textures
and experimental signatures of the resulting thermal responses remain
elusive. Here we classify the PAM textures for all 122 magnetic point
groups (MPGs) using two equivalent approaches: directly solving the linear
symmetry constraints on the multipole expansion coefficients and constructing the
allowed textures from group-representation basis functions. Both methods
yield the same fivefold classification according to the leading nonzero
order: $s$-wave (monopole), $p$-wave (dipole), $d$-wave (quadrupole),
$f$-wave (octupole), and $g$-wave (hexadecapole). We further show that a
temperature gradient generates a nonequilibrium PAM density whose $n$th-order
term in $\nabla T$ contains an energy-weighted PAM $2^n$-pole tensor and a kinematic remainder, thereby
linking the response order and partial-wave channel through exact
selection rules. Because the Berry curvature obeys the same symmetry
constraints, this taxonomy is in one-to-one correspondence with
Berry-curvature multipoles and their nonlinear Hall responses. We propose
a cantilever torque measurement whose allowed angular harmonics are fixed by the MPG, with period $2\pi/n$ for a pure $n$th harmonic, providing a direct probe of the partial-wave content of
PAM in both nonmagnetic and magnetic materials.
\end{abstract}

\maketitle

\textit{Introduction.}
Chiral phonons are lattice vibrations in which atoms trace circular or
elliptical trajectories about their equilibrium positions and thereby
carry angular momentum~\cite{mclellan1988,zhangniu2015,10.1021/acs.nanolett.4c00606,Juraschek2025a,zhang2026adv}.
Over the past decade, they have attracted broad interest in magnetism,
spintronics, valleytronics, thermal transport, and phononic
engineering~\cite{chen2019wse2,liu2019valley,du2019,delhomme2020,zhang2023gate,kim2023seebeck,luo2023rare,cui2023magnon,PRL2024}.
Chiral phonons give rise to diverse phenomena, including the phonon Hall
effect~\cite{qin2012,saito2019,Grissonnanche2020a},
temperature-gradient-induced phonon magnetization~\cite{hamada2018}, the phonon Zeeman
effect~\cite{cheng2020,baydin2022pbte,wu2023fluctuation,tang2024exciton,PRB2025,mustafa2025origin},
and  Einstein--de Haas
effect~\cite{zhangniu2014,Tauchert2022a,davies2024a,choi2024}.
Experimentally, chiral phonons have been observed in monolayer WSe$_2$,
$\alpha$-quartz, Pb$_{1-x}$Sn$_x$Te, and $\alpha$-HgS by 
infrared circular dichroism, resonant inelastic x-ray scattering, terahertz
magnetospectroscopy, and circularly polarized Raman spectroscopy,
respectively~\cite{doi:10.1126/science.aar2711,ueda2023,baydin2023,Ishito2023a}.
In magnetic materials, magneto-optical and magneto-Raman spectroscopies
have revealed chiral phonons in the ferromagnetic Weyl semimetal
Co$_3$Sn$_2$S$_2$~\cite{PhysRevLett.134.196905,che2025}.

Symmetry plays a fundamental role in determining both the classification
of chiral phonons and the responses used to detect
them~\cite{zhang2022nonsymmorphic,coh2023,gmfc-gx4v}. In a crystal,
the symmetry fixes the allowed momentum-space texture of phonon angular
momentum (PAM). For nonmagnetic crystals, Yang
\textit{et al.} classified chiral-phonon across all 230 space groups and
identify chiral phonon with higher-order helicity
patterns~\cite{yang2026a}. In magnets, however, magnetic
order breaks time-reversal symmetry $\cT$ and modifies phonons through
the molecular Berry
curvature~\cite{RevModPhys.64.51,saparov2022,ren2024prx,zhang2026mbc}.
Magnetic point groups (MPGs) therefore provide a natural symmetry
framework for classifying the allowed PAM textures in magnetic crystals.
Although classifications have been reported for collinear
magnets~\cite{wang2025altera} and two-dimensional systems~\cite{liu2026},
two central challenges remain: (i) completing the classification of PAM
textures across all 122 MPGs, and (ii) establishing the link between the
multipole structure of a PAM texture and an experimentally measurable
response.

In this Letter, we address both challenges. We first show that PAM
transforms as a time-odd axial vector [Eq.~\eqref{eq:transform}],
reducing the classification to a single linear constraint. We solve this
constraint in two independent ways: by directly solving the linear
equations for the multipole coefficients and by constructing the allowed
textures using group-representation basis
functions.
Both approaches yield identical results for all 122 MPGs. PAM vanishes
identically in the 32 groups containing space-time reversal ($\PT$) symmetry, while the remaining groups
fall into five classes according to the leading nonzero order: monopole
($s$-wave), dipole ($p$-wave), quadrupole ($d$-wave), octupole ($f$-wave),
and hexadecapole ($g$-wave). Furthermore, we show that a temperature
gradient generates a phonon magnetization at successive orders in
$\nabla T$ and identify the chiral-phonon multipole contribution at each
order. We demonstrate that the PAM texture and its thermal response obey
the same selection rules and have the same leading order, directly
linking the response order to the partial-wave channel. We analyze the
response as the in-plane temperature-gradient direction is varied and
show that symmetry fixes the allowed angular harmonics, as demonstrated
for an $f$-wave PAM texture in a breathing kagome lattice. Finally, we
propose a cantilever-torque measurement that identifies the partial-wave
channel through the angular periodicity of the
torque.
Our results establish a symmetry-based framework for classifying PAM
textures and designing experimental probes of their higher-order
multipoles in magnetic materials.

\textit{PAM and symmetry analysis.}
For a phonon branch $\nu$ at momentum $\qv$ with mass-weighted polarization
vector $\mathbf e_{\qv\nu}$, the mode angular momentum
is~\cite{zhangniu2014,zhangniu2015}
\begin{equation}
j_{\alpha}^{\qv\nu}=i\hbar\sum_{\beta\gamma}\epsilon_{\alpha\beta\gamma}\,
[\mathbf e_{\qv\nu}]^{*}_{\beta}[\mathbf e_{\qv\nu}]_{\gamma},
\label{eq:pamdef}
\end{equation}
where $\alpha=x,y,z$ labels the Cartesian component and
$\epsilon_{\alpha\beta\gamma}$ is the Levi-Civita symbol.  For a nondegenerate branch we write
$\Jt(\qv)=(j_x,j_y,j_z)$ for the resulting PAM texture. We use $g=\{\etag\Rg|\bm t\}$ to denote a magnetic
operation, in which $\Rg$ is the spatial part of
the operation and $\bm t$ is a translation; $\etag=+1$ for a unitary operation
and $\etag=-1$ for an antiunitary one, the latter including time reversal
$\cT$. The operation sends
$\qv\to\Pg\qv$ with $\Pg=\etag\Rg$, rotates the polarization vectors by
$\Rg$, and complex conjugates them when $\etag=-1$. Therefore, PAM transforms as a time-odd axial vector,
\begin{equation}
\begin{split}
\Jt(\Pg\qv)=\etag\det(\Rg)\,\Rg\,\Jt(\qv)\equiv\Ug\,\Jt(\qv).
\end{split}
\label{eq:transform}
\end{equation}
Spatial inversion and time reversal impose the constraints
$\Jt(-\qv)=\Jt(\qv)$ and $\Jt(-\qv)=-\Jt(\qv)$, respectively. Consequently, $\PT$ symmetry enforces $\Jt(\qv)=0$, ruling out nonzero PAM
in 32 of the 122 MPGs. For the remaining MPGs, the three components of $\Jt(\qv)$ are
generally nonzero at a generic momentum.

Because $\Jt(\qv)$ is analytic, it can be expanded around the $\Gamma$ point ($\bm q=0$) as
a power series in the momentum components,
\beqn
&&j_{i}(\qv)=\sum_{n=0}^{\infty}\ \sum_{j_{1},\,\dots,\,j_{n}=x,y,z}
c^{(n)}_{i|\,j_{1}\cdots j_{n}}\,q_{j_{1}}\cdots q_{j_{n}}, \nonumber\\
&&c^{(n)}_{i|\,j_{1}\cdots j_{n}}=\frac{1}{n!}
\left.\frac{\partial^{n}j_{i}}{\partial q_{j_{1}}\cdots\partial q_{j_{n}}}
\right|_{\qv=0}.
\label{eq:multipole}
\eeqn
Here the outer sum runs over the multipole order $n$ and the inner sum over each
Cartesian component $j_{a}=x,y,z$. The index $i$ is the axial (PAM) index and
$j_{1},\dots,j_{n}$ are polar (momentum) indices, the vertical bar separating the
two kinds. The integer $n$ is the multipole order---$n=0,1,2,3,4$ are the
monopole ($s$-wave), dipole ($p$-wave), quadrupole ($d$-wave), octupole ($f$-wave), and hexadecapole ($g$-wave),  denoting the
partial-wave channel.
Because $\qv\to\Pg\qv$ is linear, it preserves the total degree of each
term; terms of different degrees never mix, and
Eq.~\eqref{eq:transform} can therefore be solved order by order. Substituting Eq.~\eqref{eq:multipole} into Eq.~\eqref{eq:transform} yields
the following symmetry constraint on the multipole coefficients:
\begin{equation}
\begin{split}
c^{(n)}_{i|\,k_{1}\cdots k_{n}}={}&\etag^{\,n+1}\det(\Rg)\,[\Rg]_{ii'}\\
&\times[\Rg]_{k_{1}j_{1}}\cdots[\Rg]_{k_{n}j_{n}}
c^{(n)}_{i'|\,j_{1}\cdots j_{n}}.
\end{split}
\label{eq:tensorrule}
\end{equation}
Thus, $c^{(n)}$ is $\cT$-odd for even $n$ and $\cT$-even for odd $n$.
Moreover, $c^{(n)}$ is symmetric under permutations of its momentum
indices. For a given MPG,
Eq.~\eqref{eq:tensorrule} can be recast as a system of linear equations for
these coefficients, as detailed in the Supplemental Materials (SMs) \cite{supp}. Solving this system order by order
yields the symmetry-allowed partial-wave channels.

The symmetry-allowed partial-wave channels can also be obtained from
the Cartesian basis functions of group-representation theory. The
momentum point group that leaves the $\Gamma$ point invariant is
$\tilde G=\{\Pg\}$, with $\Pg=\etag\Rg$. For the 90 groups without $P\mathcal T$, the three components of $\Jt(\qv)$ form a
three-dimensional representation $\Gamma_{\mathrm{PAM}}$ of $\tilde G$ \cite{PhysRevX.14.031037,2025arXiv251005512L,luoxun2026},
with representation matrices $\Ug=\etag\det(\Rg)\Rg$ defined in
Eq.~\eqref{eq:transform}. The momentum dependence of $\Jt(\qv)$ is determined directly by the
basis functions of $\Gamma_{\mathrm{PAM}}$. 
Let $d_n$ count the independent degree-$n$ equivariant vector polynomials, including all representation multiplicities, and let $\mathcal N=\{n:d_n>0\}$. Using a complete vector basis, we write
\begin{equation}
\Jt(\qv)=\sum_{n\in\mathcal N}\sum_{\lambda=1}^{d_n}
a^{(n)}_{\lambda}\,\boldsymbol\Phi^{(n)}_{\lambda}(\qv),
\label{eq:irrep}
\end{equation}
where $\boldsymbol\Phi^{(n)}_{\lambda}(P_g\qv)=U_g\boldsymbol\Phi^{(n)}_{\lambda}(\qv)$ and  $a^{(n)}_{\lambda}$ is the expansion coefficient. The lowest order in
$\mathcal N$ gives the leading partial-wave channel. This construction
is equivalent to Eq.~\eqref{eq:tensorrule} and the complete enumeration are given in the Supplemental Material.

\begin{table*}[t]
\caption{Leading nonzero PAM classes among the 122 MPGs. For each class,
we list the multipole order $n$, the number of MPGs, a representative MPG
and its momentum-image group $\widetilde G$, the PAM representation
$\Gamma_{\text{PAM}}$ acting on $(j_x,j_y,j_z)$, the corresponding basis functions,
and the leading PAM texture. 
$a,b,c$ are real coefficients.}
\label{tab:taxonomy}
\centering
\fontsize{9}{11}\selectfont
\setlength{\tabcolsep}{3pt}
\setlength{\arrayrulewidth}{0.5pt}
\renewcommand{\arraystretch}{1.08}
\begin{tabular*}{\textwidth}{@{\extracolsep{\fill}}lcccllll@{}}
\hline
Leading channel & $n$ & Number & MPG & $\widetilde G$
& Representation $\Gamma_{\text{PAM}}$ & Basis functions
& Leading term $\Jt(\mathbf q)$ \\[1pt]
\hline
$s$ (monopole) & 0 & 31 & $4$ & $4$
& $A\oplus E$
& $\{1,z\}\oplus\{x,y\}$
& $(0,0,c)$ \\[1pt]
\hline
$p$ (dipole) & 1 & 39 & $2221'$ & $mmm$
& $B_{3u}\oplus B_{2u}\oplus B_{1u}$
& $\{x\}\oplus\{y\}\oplus\{z\}$
& $(a q_x,b q_y,c q_z)$ \\[1pt]
\hline
$d$ (quadrupole) & 2 & 15 & $\bar4'3m'$ & $432$
& $T_2$
& $\{yz,zx,xy\}$
& $c(q_yq_z,q_zq_x,q_xq_y)$ \\[1pt]
\hline
$f$ (octupole) & 3 & 4 & $\bar43m1'$ & $m\bar3m$
& $T_{2u}$
& \makecell[l]{
  $\{x(y^2-z^2),$\\
  $\quad y(z^2-x^2),$\\
  $\quad z(x^2-y^2)\}$}
& $c\begin{pmatrix}
  q_x(q_y^2-q_z^2)\\
  q_y(q_z^2-q_x^2)\\
  q_z(q_x^2-q_y^2)
  \end{pmatrix}$ \\[1pt]
\hline
$g$ (hexadecapole) & 4 & 1 & $m\bar3m$ & $m\bar3m$
& $T_{1g}$
& \makecell[l]{
  $\{yz(y^2-z^2),$\\
  $\quad zx(z^2-x^2),$\\
  $\quad xy(x^2-y^2)\}$}
& $c\begin{pmatrix}
  q_yq_z(q_y^2-q_z^2)\\
  q_zq_x(q_z^2-q_x^2)\\
  q_xq_y(q_x^2-q_y^2)
  \end{pmatrix}$ \\[1pt]
\hline
\end{tabular*}
\end{table*}

\textit{Multipole taxonomy.}
Using the two approaches described above, we obtain the complete set of
symmetry-allowed PAM channels for each of the 122 MPGs. Each MPG is
characterized by its leading channel, namely the symmetry-allowed channel
with the lowest multipole order $n$.  The resulting five classes are summarized in
Table~\ref{tab:taxonomy}. A monopole channel is allowed in 31 ferromagnet-like MPGs. Of these, 26
permit a monopole component along only one axis, with the other two
components first appearing at dipole order or higher. Three groups,
$2'$, $m'$, and $2/m'$, permit monopole components along two axes \cite{luo2026unifiedsymmetry},
whereas MPGs $1$ and $\bar1$ permit components along all three axes. These
groups all belong to the $s$-wave class because the classification is
determined by the lowest-order nonzero component.

We use MPG $4$ to illustrate the symmetry-allowed PAM components. With the
fourfold axis along $z$, its momentum point group is $\tilde G=C_4$, and
the three PAM components form the representation $A\oplus E$, with $j_z$
belonging to $A$ and $(j_x,j_y)$ to $E$. Here $A$ is the trivial
representation, with a constant and $q_z$ as its basis functions for
$n=0$ and $n=1$, respectively. The linear momentum functions
$(q_x,q_y)$ form a basis of $E$. Therefore, expanding each PAM component
in these basis functions gives
\beqn
j_x(\qv)&=c^{(1)}_{x|x}q_x+c^{(1)}_{x|y}q_y+O(q^2),\nonumber\\
j_y(\qv)&=-c^{(1)}_{x|y}q_x+c^{(1)}_{x|x}q_y+O(q^2),\nonumber\\
j_z(\qv)&=c^{(0)}_{z}+c^{(1)}_{z|z}q_z+O(q^2).
\label{eq:example-s}
\eeqn
Here $c^{(0)}_{z}$, $c^{(1)}_{x|x}$, $c^{(1)}_{x|y}$, and
$c^{(1)}_{z|z}$ are the expansion coefficients. The $c^{(0)}_{z}$ term
is the leading $s$-wave monopole, while all linear terms are $p$-wave
dipole contributions.

The 39 MPGs, such as in the dipole class include $222$, $mm2$, $4'$,
$31'$, $422$, and $432$. In these groups, the constant term in $\Jt(\qv)$  vanishes and the leading PAM
texture is linear in momentum. The quadrupole class comprises 15 MPGs
whose leading texture is quadratic, including $mmm$, $4/mmm$, $6/mmm$,
and $m\bar3$. The octupole class contains four MPGs, $\bar61'$, $\bar6m21'$, $\bar43m$, and $\bar43m1'$, with a leading
cubic texture, 
whereas $m\bar3m$ is the only MPG whose leading texture is quartic,
corresponding to the hexadecapole channel. Notably, these assignments refer only
to the leading nonzero order; higher-order PAM textures are also 
allowed. In the SMs \cite{supp}, we present a complete list of the 90 MPGs that support nonzero PAM, together with
their leading-order classifications, momentum point groups $\tilde G$,
and the three-dimensional representations formed by $\Jt(\qv)$.

\textit{Higher-order thermal transport.}
We now show that a temperature gradient generates a nonequilibrium PAM density,
with the $n$th-order contribution in $\nabla T$ containing an energy-weighted PAM $2^n$-pole moment and a kinematic remainder. We evaluate this response for a phonon mode
$(\qv,\nu)$ with frequency $\omega_{\qv\nu}$, group velocity
$\vv_{\qv\nu}=\partial\omega_{\qv\nu}/\partial\qv$, and mode PAM
$\bm{j}_{\qv\nu}$. PAM density and magnetization generally require a mode-dependent gyromagnetic conversion; here we calculate the angular-momentum density directly. Its occupation is written as
$f_{\qv\nu}=f_0+\delta f_{\qv\nu}$, where
$f_0=(e^{\hbar\omega_{\qv\nu}/k_{\mathrm B}T}-1)^{-1}$ is the
equilibrium Bose distribution at temperature $T$, and
$\delta f_{\qv\nu}$ is the nonequilibrium correction under a temperature
gradient. The PAM density is then
$j_i=(1/V)\sum_{\qv\nu}j_{\qv\nu,i}f_{\qv\nu}$, with $V$ the volume.
In MPGs whose leading allowed texture has degree $n>0$, the equilibrium contribution
$(1/V)\sum_{\qv\nu}f_0j_{\qv\nu,i}$ vanishes, whereas the
nonequilibrium correction can produce a nonzero PAM density through
$(1/V)\sum_{\qv\nu}j_{\qv\nu,i}\delta f_{\qv\nu}$, analogous to the Edelstein effect in electronic systems \cite{EDELSTEIN1990233}. Note that the linear case ($n=1$) has been studied theoretically~\cite{hamada2018}, and demonstrated experimentally~\cite{zhang2025}. We therefore focus on the
generalization to higher multipole channels and higher orders in the
temperature gradient.

In the relaxation-time approximation, the steady-state Boltzmann
equation reads
$\vv_{\qv\nu}\cdot\nabla_{\mathbf r}f_{\qv\nu}
=-\delta f_{\qv\nu}/\tau$, where $\tau$ is the phonon relaxation time. Using
$\nabla_{\mathbf r}f_0=\mathbf G\,\partial_Tf_0$ with
$\mathbf G=\nabla T$, and collecting terms with the same power of
$\mathbf G$, we obtain the iterative equation $\delta f^{(n)}_{\qv\nu}
=-\tau\,\vv_{\qv\nu}\cdot\nabla_{\mathbf r}\,
\delta f^{(n-1)}_{\qv\nu}$, where $\delta f^{(0)}_{\qv\nu}=f_0 $ and $ n\geq1$.
For a spatially uniform temperature gradient and a mode-independent
$\tau$, iterating this equation gives the $n$th-order corrections to
the nonequilibrium distribution and the induced PAM:
\begin{align}
\delta f^{(n)}_{\qv\nu}
&=(-\tau)^n(\vv_{\qv\nu}\cdot\mathbf G)^n
\partial_T^n f_0,\nonumber\\
\delta j_i^{(n)}
&=\frac{(-\tau)^n}{V}\sum_{\qv\nu}
j_{\qv\nu,i}(\vv_{\qv\nu}\cdot\mathbf G)^n
\partial_T^n f_0 .
\label{eq:nthresponse}
\end{align}
Thus, the total induced PAM is
$\delta j_i=\sum_{n\geq1}\delta j_i^{(n)}$.

To reveal the multipole structure, we use
$\partial_Tf_0=-(\omega/T)\partial_\omega f_0$ and convert each velocity
factor into a momentum derivative through
$(\mathbf G\cdot\vv)\partial_\omega f_0
=(\mathbf G\cdot\nabla_{\qv})f_0$.
Repeated application of these identities, followed by successive
integration by parts over the Brillouin zone, yields the general
$n$th-order response
\begin{equation}
\begin{split}
\delta j_i^{(n)}
={}&\left(-\frac{\tau}{T}\right)^n
\Bigl(\mathcal M^{(n)}_{i|j_1\cdots j_n}
+\mathcal M^{(n,\mathrm{kin})}_{i|j_1\cdots j_n}\Bigr)\\
&\times G_{j_1}\cdots G_{j_n},
\qquad n\geq1 ,
\end{split}
\label{eq:Jgeneral}
\end{equation}
where the central energy-weighted PAM multipole is
\begin{equation}
\mathcal M^{(n)}_{i|j_1\cdots j_n}
=\frac1V\sum_{\qv\nu}f_0\,
\partial_{q_{j_1}}\!\cdots\partial_{q_{j_n}}
\left(\omega_{\qv\nu}^{\,n}j_{\qv\nu,i}\right).
\label{eq:PAMgeneral}
\end{equation}
The kinematic remainder $\mathcal M^{(n,\mathrm{kin})}$ vanishes for $n=1$. At higher order it contains both frequency-weighting and dispersion-derivative terms and need not be small. In the SMs \cite{supp}, we give the exact expressions through fourth order and the general recurrence.

Explicitly, the first three responses are
\beqn
&&\delta j_i^{(1)}
=-\frac{\tau}{T}\,\mathcal D_{i|j}G_j ,
\label{eq:J1}\nonumber\\
&&\delta j_i^{(2)}
=\left(\frac{\tau}{T}\right)^2
\left(\mathcal Q_{i|jk}+\mathcal Q^{(\mathrm{kin})}_{i|jk}\right)G_jG_k ,
\label{eq:J2}\nonumber\\
&&\delta j_i^{(3)}
=-\left(\frac{\tau}{T}\right)^3
\left(\mathcal O_{i|jkl}+\mathcal O^{(\mathrm{kin})}_{i|jkl}\right)
G_jG_kG_l ,
\label{eq:J3}
\eeqn
where $\mathcal M^{(1)}=\mathcal D$,
$\mathcal M^{(2)}=\mathcal Q$, and $\mathcal M^{(3)}=\mathcal O$, with
\beqn
&&\mathcal D_{i|j}
=\frac1V\sum_{\qv\nu}f_0\,
\partial_{q_j}\!\left(\omega_{\qv\nu}j_{\qv\nu,i}\right),
\label{eq:PAMdipole}\nonumber\\
&&\mathcal Q_{i|jk}
=\frac1V\sum_{\qv\nu}f_0\,
\partial_{q_j}\partial_{q_k}\!
\left(\omega_{\qv\nu}^{2}j_{\qv\nu,i}\right),
\label{eq:PAMquadrupole}\nonumber\\
&&\mathcal O_{i|jkl}
=\frac1V\sum_{\qv\nu}f_0\,
\partial_{q_j}\partial_{q_k}\partial_{q_l}\!
\left(\omega_{\qv\nu}^{3}j_{\qv\nu,i}\right).
\label{eq:PAMoctupole}
\eeqn
These are respectively the dipole, quadrupole, and octupole moments of
the energy-weighted PAM textures $\omega\bm{j}$,
$\omega^2\bm{j}$, and $\omega^3\bm{j}$. The kinematic-remainder terms
affect the response magnitude but obey exactly the same symmetry
constraints as the corresponding PAM multipole.

Expanding the derivatives makes the physical content explicit. For
example,
\begin{equation}
\mathcal D_{i|j}
=\frac1V\sum_{\qv\nu}f_0
\left(v_{\qv\nu,j}j_{\qv\nu,i}
+\omega_{\qv\nu}\,\partial_{q_j}j_{\qv\nu,i}\right),
\end{equation}
which contains both the velocity-weighted PAM and the dipole of the PAM
texture itself. The tensors $\mathcal Q$ and $\mathcal O$ analogously
contain second and third derivatives of the PAM texture. Together with
the monopole $M_i$ ($s$ wave), they form the PAM multipole ladder:
$\mathcal D$ ($p$ wave), $\mathcal Q$ ($d$ wave), and $\mathcal O$
($f$ wave). More generally, $\mathcal M^{(n)}$ and its kinematic-remainder
counterpart transform exactly as the degree-$n$ PAM texture coefficient
$c^{(n)}$ in Eq.~\eqref{eq:tensorrule}: each behaves as an axial
rank-$(n+1)$ tensor, symmetric in its last $n$ indices, with
time-reversal parity $(-1)^{n+1}$. The thermal response therefore obeys
exactly the same MPG selection rules as the corresponding PAM multipole:
the $n$th-order response is symmetry-allowed precisely when a degree-$n$ PAM texture is allowed (Table~\ref{tab:rules}).

\begin{table}[t]
\caption{\label{tab:rules}Symmetry selection rules for PAM partial-wave
channels and thermal responses. Counts include all MPGs allowing each
channel, not only those with it as the leading channel.}
\centering
\fontsize{9}{11}\selectfont
\setlength{\tabcolsep}{4pt}
\renewcommand{\arraystretch}{1.2}
\begin{tabular}{cccl}
\hline
Channel & \makecell{Allowed\\MPGs} & \makecell{Multipole\\tensor}
& PAM observable\\
\hline
$s$ & 31 & $M$ & equilibrium density\\
$p$ & 58 & $\mathcal{D}$ & linear thermal response\\
$d$ & 66 & $\mathcal{Q}$ & quadratic thermal response\\
$f$ & 69 & $\mathcal{O}$ & cubic thermal response\\
$g$ & 69 & $\mathcal{M}^{(4)}$ & quartic thermal response\\
\hline
\end{tabular}
\end{table}

\begin{figure}[t]
\centering
\includegraphics[width=1\columnwidth]{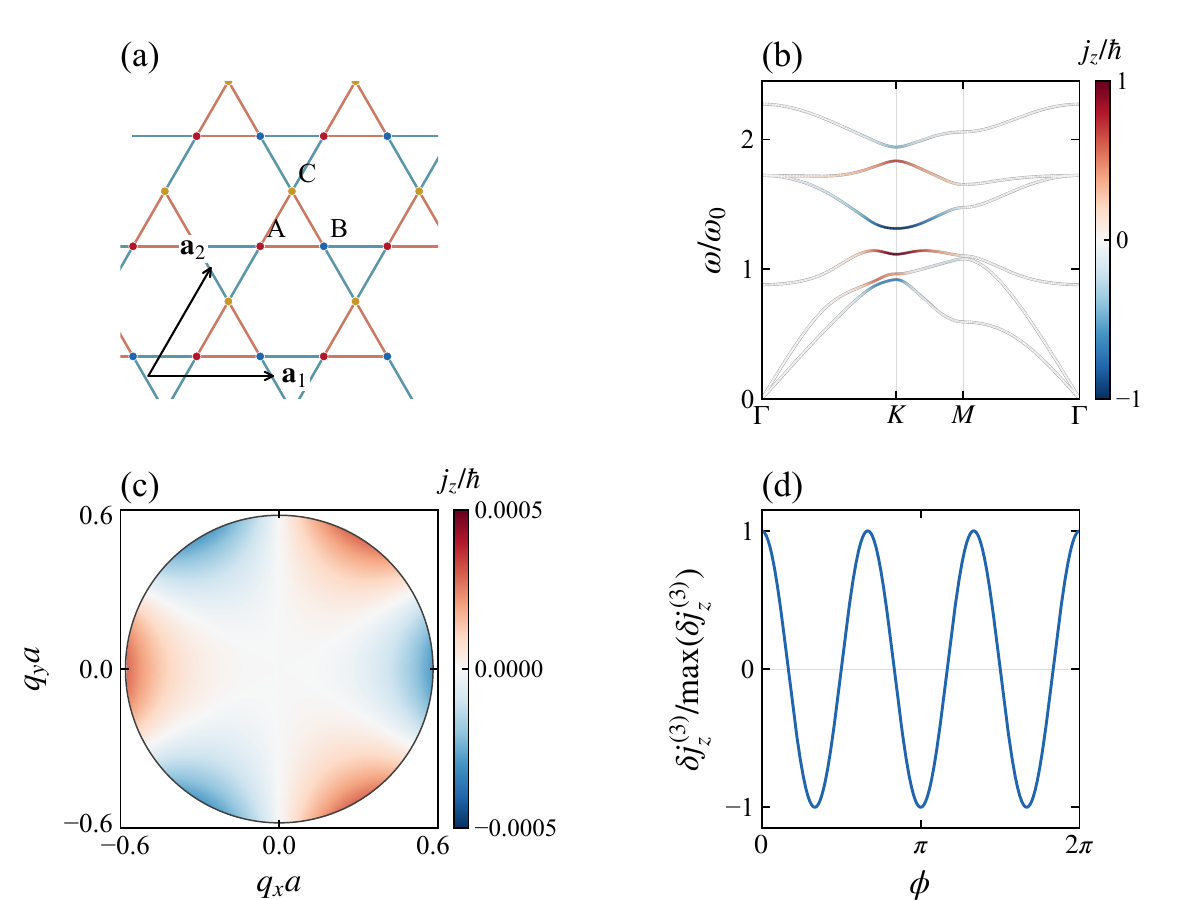}
\caption{Breathing-kagome model and its $f$-wave PAM response.
(a) Three-site lattice with primitive vectors $\mathbf a_1$ and
$\mathbf a_2$ and distinct intra- and intercell bonds.
(b) Six phonon branches colored by the mode-resolved $l_z/\hbar$.
(c) Momentum-space $l_z$ of the highest optical branch within
$|q|a\leq0.6$, showing six lobes of alternating sign.
(d) Normalized third-order PAM density response $\delta j_z^{(3)}$ to an
in-plane temperature gradient. The response follows
$\cos3\phi$ and has period $2\pi/3$.}
\label{fig2}
\end{figure}

\textit{Breathing-kagome lattice model.}
We use a two-dimensional breathing kagome lattice
[Fig.~\ref{fig2}(a)]~\cite{chen2019} to realize a leading $f$-wave PAM
texture. The unequal bond strengths on the upward and downward triangles
break inversion while preserving $C_{3v}$ and time reversal, yielding the
gray magnetic point group $3m1'$. In two-dimensional momentum space, $j_z$
transforms as the $A_2$ representation of $C_{3v}$ and is odd under time
reversal. The lowest allowed term is therefore cubic,
\begin{equation}
j_z(\qv)
\propto c_f\left(q_x^3-3q_xq_y^2\right)+O(q^5),
\label{eq:kagome-leading}
\end{equation}
corresponding to an $f$-wave octupole texture.

The six phonon bands comprise two acoustic and four optical branches
[Fig.~\ref{fig2}(b)].  We focus on the highest optical branch, which is
nondegenerate at $\Gamma$. Its calculated PAM exhibits six lobes of
alternating sign [Fig.~\ref{fig2}(c)], confirming the $f$-wave texture.
For the thermal PAM-density response, the $\mathcal D$ and $\mathcal Q$
channels are forbidden by symmetry in this two-dimensional gray group.
The leading contribution is therefore third order and arises through the
$\mathcal O$ channel. Evaluating the full response using all six branches
gives $\delta j_z^{(3)}(\phi)\propto\cos3\phi$ [Fig.~\ref{fig2}(d)], where
$\phi$ is the azimuthal angle of the in-plane temperature gradient. The
resulting threefold angular oscillation is symmetry enforced \cite{supp}. The model and numerical details are given in the SMs~\cite{supp}, where we also classify the PAM textures allowed by
all two-dimensional MPGs and identify six leading-order
partial waves: the $s$, $p$, $d$, $f$, $g$, and $i$ waves.

\textit{Discussion and Summary.} The symmetry-fixed periodicity of the PAM response offers a direct
experimental route to identifying the partial-wave channels. In the
SMs \cite{supp}, we systematically analyze the response of PAM
multipoles to in-plane thermal gradients applied along different directions
and determine, for each MPG, the angular periodicity of
$\delta j$ as the gradient is rotated. As the gradient-induced phonon angular
momentum relaxes, its transfer to the lattice generates a mechanical torque
that can be measured using a microcantilever \cite{zhang2025,chiralometer}. The direction of the thermal
gradient can be controlled by translating a focused laser-heating spot
relative to the sample, allowing the torque to be recorded as a function of
the gradient angle. The measured angular periodicity thus provides an
experimental fingerprint of the partial-wave channel.

Notably, our classification of PAM multipoles is closely connected to
Berry-curvature multipoles and the nonlinear Hall effect in electronic
systems \cite{PhysRevB.107.115142}. Although they have distinct microscopic origins, the $n$th PAM and
Berry-curvature multipoles both transform as rank-$(n+1)$ axial tensors and
therefore obey exactly the same symmetry selection rules. Consequently,
nonlinear Hall materials motivate searches for symmetry-allowed PAM responses. The corresponding electronic Berry-curvature mechanism is order $n+1$ in electric field, whereas the PAM-density response here is order $n$ in temperature gradient.

In summary, we have classified momentum-space PAM textures for all 122
MPGs. MPGs allowing nonzero PAM fall into five classes
according to their leading multipole order: $s$ (monopole), $p$ (dipole),
$d$ (quadrupole), $f$ (octupole), and $g$ (hexadecapole) waves. The
$n$th-order thermal-gradient-induced PAM density is governed by the
energy-weighted PAM $2^{n}$-pole tensor together with its kinematic remainder, giving identical symmetry selection rules
for the PAM texture and thermal response. We also extend the framework to
two-dimensional MPGs and propose cantilever-based torque measurements to
identify partial-wave channels through their angular periodicity.

\medskip
\noindent
\textit{Note added.}---During the preparation of this work, we became aware of
a related study~\cite{liu2026}, which classifies PAM in two-dimensional
magnets and investigates the associated thermal transport.

\textit{Acknowledgments.}---Xun-Jiang Luo thanks Ding-Fu Shao and Rui-Chun Xiao for helpful
discussions. This work was supported by the  the National
Natural Science Foundation of China (Grant Nos. 12604254).

\bibliography{reference}

\end{document}